\documentclass[longauth]{aa}
\usepackage{natbib}
\bibpunct{(}{)}{;}{a}{}{,}% to follow the A&A style

\usepackage{xcolor}
\usepackage{ulem}
\usepackage{amsmath}
\usepackage{amsfonts}
\usepackage{amssymb}
\usepackage[switch]{lineno}
\usepackage{booktabs}
\usepackage{float}

\usepackage[varg]{txfonts}

\usepackage{graphicx}
\usepackage{multirow}
 
\usepackage{hyperref}
\hypersetup{
    colorlinks=true,
    linkcolor=blue,
    citecolor=blue,
    urlcolor=blue
}

\begin{document} 

%\linenumbers

\title{Detection of the Geminga pulsar at energies down to 20 GeV with the LST-1 of CTAO}

\author{
K.~Abe\inst{1} \and
S.~Abe\inst{2} \and
A.~Abhishek\inst{3} \and
F.~Acero\inst{4,5} \and
A.~Aguasca-Cabot\inst{6} \and
I.~Agudo\inst{7} \and
C.~Alispach\inst{8} \and
D.~Ambrosino\inst{9} \and
F.~Ambrosino\inst{10} \and
L.~A.~Antonelli\inst{10}
% ...
}
\institute{
Department of Physics, Tokai University, 4-1-1, Kita-Kaname, Hiratsuka, Kanagawa 259-1292, Japan
\and Institute for Cosmic Ray Research, University of Tokyo, 5-1-5, Kashiwa-no-ha, Kashiwa, Chiba 277-8582, Japan
\and INFN and Università degli Studi di Siena, Dipartimento di Scienze Fisiche, della Terra e dell'Ambiente (DSFTA), Sezione di Fisica, Via Roma 56, 53100 Siena, Italy
\and Université Paris-Saclay, Université Paris Cité, CEA, CNRS, AIM, F-91191 Gif-sur-Yvette Cedex, France
\and FSLAC IRL 2009, CNRS/IAC, La Laguna, Tenerife, Spain
\and Departament de Física Quàntica i Astrofísica, Institut de Ciències del Cosmos, Universitat de Barcelona, IEEC-UB, Martí i Franquès, 1, 08028, Barcelona, Spain
\and Instituto de Astrofísica de Andalucía-CSIC, Glorieta de la Astronomía s/n, 18008, Granada, Spain
\and Department of Astronomy, University of Geneva, Chemin d'Ecogia 16, CH-1290 Versoix, Switzerland
\and INFN Sezione di Napoli, Via Cintia, ed. G, 80126 Napoli, Italy
\and INAF - Osservatorio Astronomico di Roma, Via di Frascati 33, 00040, Monteporzio Catone, Italy
% ...
}

%%%%%%%%%%%%%%%%%%%%%%%%%

\date{Received 3 March 2025 / Accepted 21 May 2025}

\abstract
{
    Geminga is the third gamma-ray pulsar firmly detected by imaging atmospheric Cherenkov telescopes (IACTs) after the Crab and the Vela pulsars. Most of its emission is expected at tens of giga-electronvolts, and, out of the planned telescopes of the upcoming Cherenkov Telescope Array Observatory (CTAO), the Large-Sized Telescopes (LSTs) are the only ones with optimised sensitivity at these energies.
}
{
    We aim to characterise the gamma-ray pulse shape and spectrum of Geminga as observed by the first LST (hereafter LST-1) of the Northern Array of CTAO. Furthermore, this study confirms the great performance and the improved energy threshold of the telescope, as low as 10 GeV for pulsar analysis, with respect to current-generation Cherenkov telescopes.    
}
{
    We analysed 60 hours of good-quality data taken by the LST-1 between December 2022 and March 2024 at zenith angles below 50°. Additionally, a new \textit{Fermi}-LAT analysis of 16.6 years of data was carried out to extend the spectral analysis down to 100 MeV. Lastly, a detailed study of the systematic effects was performed. 
}
{
    We report the detection of Geminga in the energy range between 20 and 65 GeV. Of the two peaks of the phaseogram, the second one, P2, is detected with a significance of 12.2$\sigma$, while the first (P1) reaches a significance level of 2.6$\sigma$. The best-fit model for the spectrum of P2 was found to be a power law with a spectral index of $\Gamma=(4.5\pm0.4_{stat})^{+0.2_{sys}}_{-0.6_{sys}}$, compatible with the previous results obtained by the MAGIC Collaboration. No evidence of curvature is found in the LST-1 energy range. The joint fit with \textit{Fermi}-LAT data confirms a preference for a sub-exponential cut-off over a pure exponential, even though both models fail to reproduce the data above several tens of giga-electronvolts. The overall results presented in this paper prove that the LST-1 is an excellent telescope for the observation of pulsars, and improved sensitivity is expected to be achieved with the full CTAO Northern Array.
}
{}


\titlerunning{The Geminga pulsar with LST-1}
\authorrunning{CTAO-LST Project}
\maketitle

\section{Introduction}
Pulsars make up the majority of the Galactic gamma-ray sky, with almost 340 identified sources between rotation-powered and millisecond pulsars in the Third \textit{Fermi} Large Area Telescope (LAT) catalogue of gamma-ray pulsars, 3PC \citep{Smith_2023}. Their gamma-ray emission is usually explained with curvature radiation models, which are compatible with the double-peaked phase-folded light curves and the sub-exponential cut-offs at a few giga-electronvolts observed in the spectra of \textit{Fermi}-LAT pulsars. Imaging atmospheric Cherenkov telescopes (IACTs) later detected emission above 25 GeV from the Crab pulsar \citep{MAGIC_Crab_25gev}. Then, more surprisingly, the Crab was discovered to show pulsed emission above 100 GeV \citep{Crab_veritas_2011, Crab_magic_2012} and even up to 1.5 TeV \citep{Ansoldi_2016}. More recently, \cite{Vela_2023} discovered pulses up to 20 TeV from Vela. The detection of tera-electronvolt pulsed emission challenges the capability of curvature radiation models in explaining the entire gamma-ray spectrum from mega-electronvolt to tera-electronvolt energies.

Despite the important achievement, current-generation IACTs have limited sensitivity for the detection of pulsars. The Cherenkov Telescope Array Observatory (CTAO) \citep{ctao} will be the next-generation facility for ground-based gamma-ray observations, consisting of two arrays, one in the Northern Hemisphere and one in the Southern, composed of telescopes of three different sizes. This configuration will allow it to cover a wide energy range from around 20 GeV to almost 300 TeV. In particular, Large-Sized Telescopes (LSTs) are designed to have the highest sensitivity in the energy range from 20 to 200 GeV \citep{lst_icrc_2023}, being ideal instruments for studying pulsars in the gap between \textit{Fermi}-LAT and the tera-electronvolt domain, covered by other IACT arrays. The first LST of the CTAO-North array, dubbed LST-1, has already been built and is currently in the final commissioning phase.

Geminga (PSR J0633+1746) is the third pulsar detected above 5$\sigma$ significance by IACTs\footnote{Note that a fourth pulsar, PSR B1706-44, was reported with a 4.7$\sigma$ significance by the H.E.S.S. Collaboration \citep{psr_b1706}.} \citep{Geminga_magic}. It is a well-known and studied source, considered the archetype of a middle-aged radio-quiet gamma-ray pulsar. It is also one of the closest pulsars to us, with an estimated parallax distance of $d=250^{+120}_{-62}$ pc \citep{geminga-dist}. The SAS-2 satellite discovered it as an isolated gamma-ray source in 1972 without any counterpart at other wavelengths \citep{geminga_sas2}. Only twenty years later it was possible to identify it as a pulsar thanks to combined results from X-ray observations by ROSAT \citep{geminga_rosat} and gamma-ray data from EGRET \citep{geminga_egret}. From these last results, the main rotation parameters were estimated to be a period of $P=237$ ms and a period derivative of $\dot{P} \simeq 1.1 \cdot 10^{-14}$ s s$^{-1}$, from which they derived its characteristic age of $\tau_c \sim 300$ kyr and its spin-down power $\dot{E} \sim 3.5 \cdot 10^{34}$ erg s$^{-1}$. There is still no detection of Geminga at radio wavelengths, and the latest results report an upper limit on the average flux density of 0.4 - 4 mJy at a frequency of 111 MHz \citep{geminga_radio}, where the range is due to different assumptions on the pulse duration. Despite being radio-quiet, Geminga is one of the few pulsars also detected at optical frequencies \citep{optical_pulsars}.

Geminga was detected within one year of operations by \textit{Fermi}-LAT \citep{geminga_fermi}, and, more surprisingly, emission above 15 GeV was reported by \cite{Geminga_magic}. They detected the second emission peak of the pulsar at a significance level of 6.3$\sigma$ and fitted its spectrum using a power law (PL) with spectral index $\Gamma=5.62 \pm 0.54$ extending up to 75 GeV. This could be interpreted as the smooth transition from a regime dominated by curvature radiation to an inverse-Compton-dominated regime at higher energies.

In this paper, we report the detection of Geminga with the LST-1. Section \ref{sec:obs} summarises the details of the data analysis; in Sect. \ref{sec:results} we present the main results obtained in this work, in Sect. \ref{sec:syst} we outline the tests we performed to study systematic effects on the LST-1 analysis; finally, in Sect. \ref{sec:final} we discuss the results and present our conclusions.

\section{Observations and data analysis} \label{sec:obs}
\subsection{LST-1 observations and data analysis}
LST-1 observations were performed in 20-minute-long runs in the so-called wobble mode \citep{fomin_94}, with an off-set angle of 0.4° from the camera centre. The LST-1 observed Geminga as part of the commissioning program between December 2022 and March 2024 for almost 80 hours in total. For this analysis, we required observations in dark conditions only, i.e. no dusk and down or moonlight, and performed a quality selection of the data to reject the runs affected by telescope problems or bad weather conditions. The method chosen to assess the data quality is based on the analysis of the differential intensity spectra of the detected showers, which are mostly induced by hadrons. The spectrum is characterised by a peak followed by a power law decay. The method consists of fitting the power-law region of the spectrum, then the selection is based on a range of values derived empirically from a good sample of observations. This approach is an evolution of the one used in \cite{performance_paper}. The final sample was reduced to 60 hours of good-quality data taken at zenith distance (\textit{Zd}) lower than 50°, among which the majority (54 hours) were taken at \textit{Zd} < 25°, which helps to reduce the overall energy threshold in the analysis.

The data were then reduced using version \texttt{v0.10.7} of the \texttt{cta-lstchain} software library \citep{lstchain_adass_2020, lstchain_software}, following the procedure described in \cite{performance_paper}. Three main characteristics of the primary particle that generates the shower were reconstructed: its energy, incoming direction in the sky, and nature. The latter was measured with the `gammaness' parameter, an indicator of how likely it is that the primary particle is a hadron (gammaness $\sim$ 0) or a gamma ray (gammaness $\sim$ 1). The reconstruction is performed using random forest methods trained on simulated Monte Carlo (MC) data. The package \texttt{lstmcpipe} \citep{garcia2022lstmcpipe, lstmcpipe_software}, which uses \texttt{lstchain}, was employed to process the simulated data. In particular, an MC production of gamma rays and protons was simulated along a declination line of $22.76^\circ$ as seen from the LST-1 site, close to the declination of Geminga in the sky. We also tuned the night sky background in the simulations to match the level observed in the sky region around Geminga. The instrument response functions (IRFs) of the LST-1 were computed using an MC test sample simulated in different sky positions along the declination line, and they were subsequently interpolated with the telescope pointing to obtain the correct IRFs for each observation run. 

We adopted the source-dependent approach, for which the source position is assumed a priori. In this way, we can achieve a better performance in the reconstruction of low-energy showers \citep{performance_paper}. The main parameter introduced when performing a source-dependent analysis is `alpha', defined as the angle between the main axis of the shower ellipse and the line connecting the centroid of the image and the position of the source.

To achieve a good agreement between the simulations and the observed data, especially for the weaker signals, we applied to both samples a cut on the image intensity of $I > 50$ photoelectrons (p.e.). This cut allowed us also to reject the lowest intensity events, which are usually poorly reconstructed. To remove the hadronic background, we also applied energy-dependent cuts on alpha and gammaness based on a standard 70\% selection efficiency of the MC. Later, in Sect. \ref{sec:syst}, we study the systematic effects on the analysis when varying these cuts.

The next step of the analysis chain was the computation of the rotational phase of the pulsar at which each event was emitted. This was done using version \texttt{v0.9.7} of the \texttt{PINT} Python package \citep{pint_software, pint_article} through the \texttt{PulsarTimingAnalysis} software \citep{pulsar_software}. The pulsar's ephemeris were computed based on \textit{Fermi}-LAT data starting on 6 August 2008 and available online\footnote{\url{https://www.mpp.mpg.de/~ceribell/geminga/index.php}}. Finally, the science products, i.e. the phaseogram and the spectral energy distribution (SED), were obtained using version \texttt{v1.1} of \texttt{Gammapy} \citep{gammapy_2023, gammapy_software}.

The phase regions adopted for this analysis are the ones defined in \cite{Geminga_magic}: P1 = [0.056; 0.161] for the first peak, P2 = [0.550; 0.642] for the second, and OFF = [0.700; 0.950] for the background phase region. We also considered the inter-peak (or Bridge) emission, defined as Bridge = [0.161; 0.550].

\subsection{\textit{Fermi}-LAT observations and data analysis} \label{sec:fermi}
Most of the gamma-ray emission from Geminga is concentrated below 10 GeV, where \textit{Fermi}-LAT is more sensitive than the LST-1. To expand the spectrum over a broader energy range, we analysed 16.6 years of public \textit{Fermi}-LAT data, starting on 4 August 2008 and lasting until 27 March 2025.

\textit{Fermi}-LAT data were reduced and analysed using version \texttt{v1.3.1} of the \textit{Fermipy} package \citep{fermipy} with version \texttt{v2.2.0} of the \textit{Fermi} Science Tools and the \texttt{P8R3\_SOURCE\_V3} IRFs. We chose a region of interest (ROI) with a side of 15$^{\circ}$ centred on the Geminga pulsar position (RA = 06\textsuperscript{h} 33\textsuperscript{m} 54.15\textsuperscript{s}, DEC = +17° 46' 12.91'') and binned in 0.1$^\circ$ bins, and the energy range 100 MeV - 300 GeV. We filtered for good timing and nominal state of the detector. For energies below $10^{3.5} \sim 3.2$ GeV, we selected the events classified as \texttt{SOURCE} (event class 128) and type 32 (\texttt{PSF3}, the best quartile of the PSF distribution), with a maximum zenith distance of 90$^\circ$. For energies above $\sim3.2$ GeV, instead, we selected the \texttt{SOURCE} class events, but we included all event types 4, 8, 16, 32 (\texttt{PSF0 + PSF1 + PSF2 + PSF3}, corresponding to no filtering for the PSF quality), with a maximum zenith distance of 105$^\circ$. This event selection allows us to limit systematics related to the PSF modelling and affecting a bright source such as Geminga at the lowest energies, while keeping sound statistics at the higher end of the spectrum. 

The last step was to obtain the spectrum and the flux points for the two peaks, for which we selected the same phase regions as for the LST-1 analysis. We performed a binned likelihood analysis in the range 100 MeV - 300 GeV with eight logarithmically spaced energy bins per decade. Then, we used a spectral-spatial model considering the sources reported in the 4FGL catalogue \citep{Abdollahi_2020} and within a region of 20$^\circ$. We included in the fit the Galactic (\texttt{gll\_iem\_v07.fits}) and isotropic background (\texttt{iso\_P8R3\_SOURCE\_V3\_v1.txt}) models, specifying the latter according to each different PSF class separately. We first ran an optimisation to estimate the predicted count rates of the sources in the sky region and froze all the parameters to their catalogue values for sources with predicted count values of less than one. We then modelled the Geminga spectrum as a power law with sub-exponential cut-off (PLSEC), using the \texttt{PLSuperExpCutoff2} model, first with free normalisation, spectral index and exponential factor, but leaving the exponential index at the 4FGL catalogue value. We also let free the normalisation of the background sources within 10$^\circ$ and with at least a 3$\sigma$ significance, as well as the shape parameters of the background sources within 7$^\circ$ with at least a 5$\sigma$ significance, and the spectral parameters of the second brightest source in the ROI, 4FGL J0617.2+2234e. Starting from the maximum likelihood values of the previous fit, we performed a further step by freezing background sources, but freeing all Geminga parameters, including the exponential index. Finally, we obtained the Geminga SED points by fitting a power law with spectral index $\Gamma=2$ in each energy bin, with the background sources locked.

\section{Results} \label{sec:results}
\subsection{Phaseogram} \label{sec:phaseogram}
\begin{figure}[h]
    \resizebox{\hsize}{!}{\includegraphics{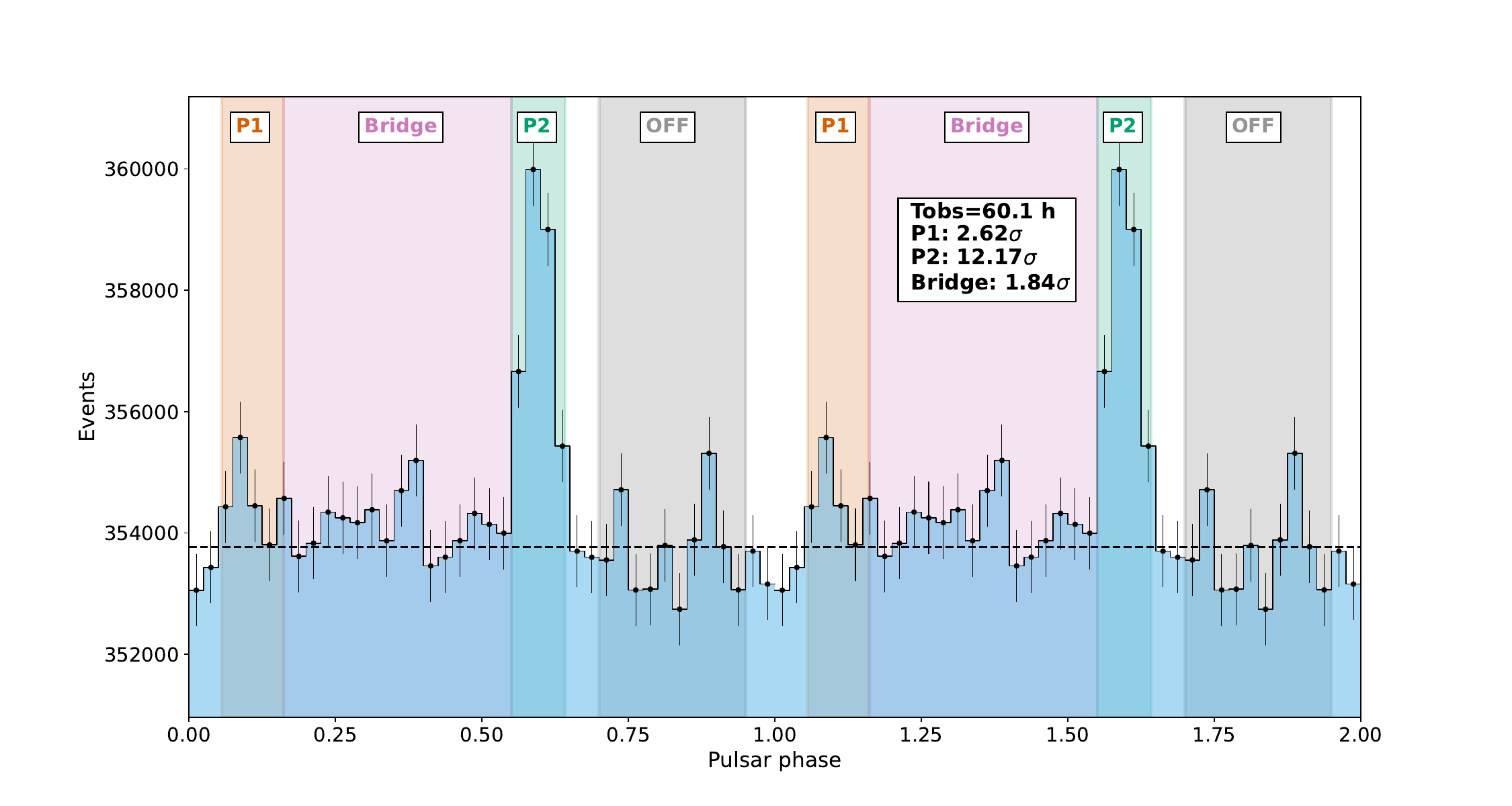}}
    \caption{Phaseogram of the LST-1 observations of the Geminga shown over two rotational periods, with no cut in energy. The different phase regions (P1, P2, Bridge and background, or OFF) are highlighted in the plot. The average level of the background counts is reported as the horizontal dashed line. We also show the Li\&Ma significance of both peaks and the inter-peak region and the total observation time.}
    \label{fig:phaseogram}
\end{figure}

Figure \ref{fig:phaseogram} depicts the phaseogram of Geminga as observed by the LST-1 at \textit{Zd} < 50$^\circ$ without applying any energy cut to the sample. Out of the two well-known peaks, only P2 is detected with a Li\&Ma significance \citep{li_and_ma} of 12.2$\sigma$. P1 reaches a significance level of 2.6$\sigma$. We also studied the signal in the Bridge region and observed an excess corresponding to a Li\&Ma significance of $1.8\sigma$.

We expect that the significance evolves as the square root of the observational time, i.e. $\sigma = A\sqrt{t(\mathrm{h})}$, so we used this relation to derive the coefficient $A$ and estimate how many hours would be needed to reach a 5$\sigma$ significance of P1. The test yielded $\sigma_{P1}=(0.34 \pm 0.13) \  t(\mathrm{h})^{1/2}$, where the error on $A$ has been estimated considering that the Li\&Ma significance, in the case of weak sources, has an uncertainty of one by definition \citep{li_and_ma}. Considering this trend, the estimated observational time for a 5$\sigma$ detection of P1 would be more than 200 hours, which could be achieved with few observational cycles of LST-1. The result for P2 proves the excellent detection capabilities of the LST-1 compared to other existing IACTs: we obtained a doubled significance when compared to \cite{Geminga_magic} in less observational time and with a single telescope.

\begin{figure*}
    \centering
    \includegraphics[width=0.4\linewidth]{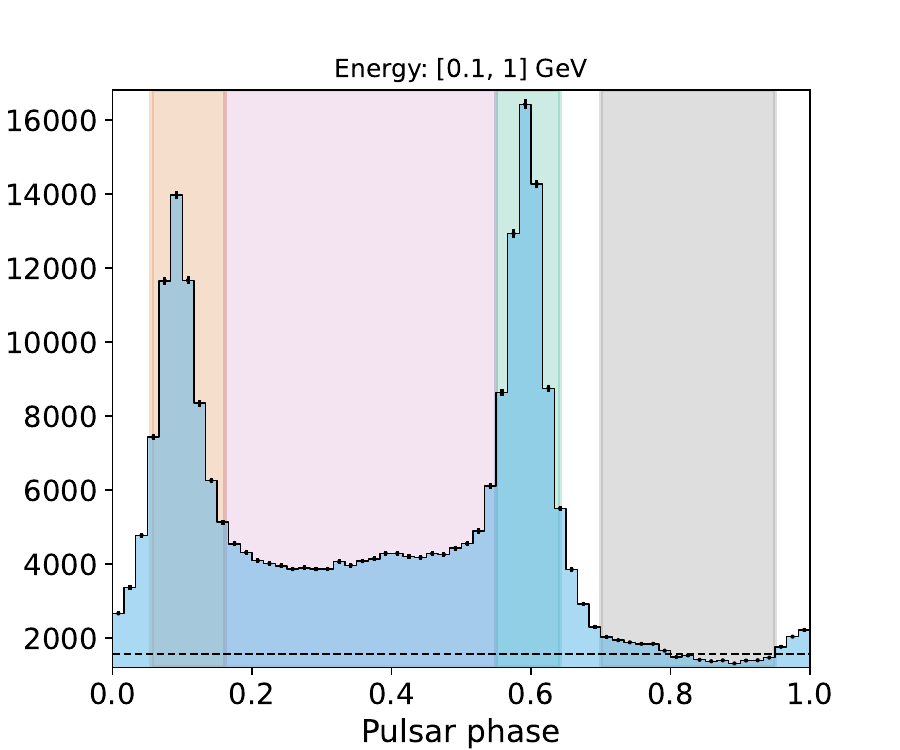}
    \includegraphics[width=0.4\linewidth]{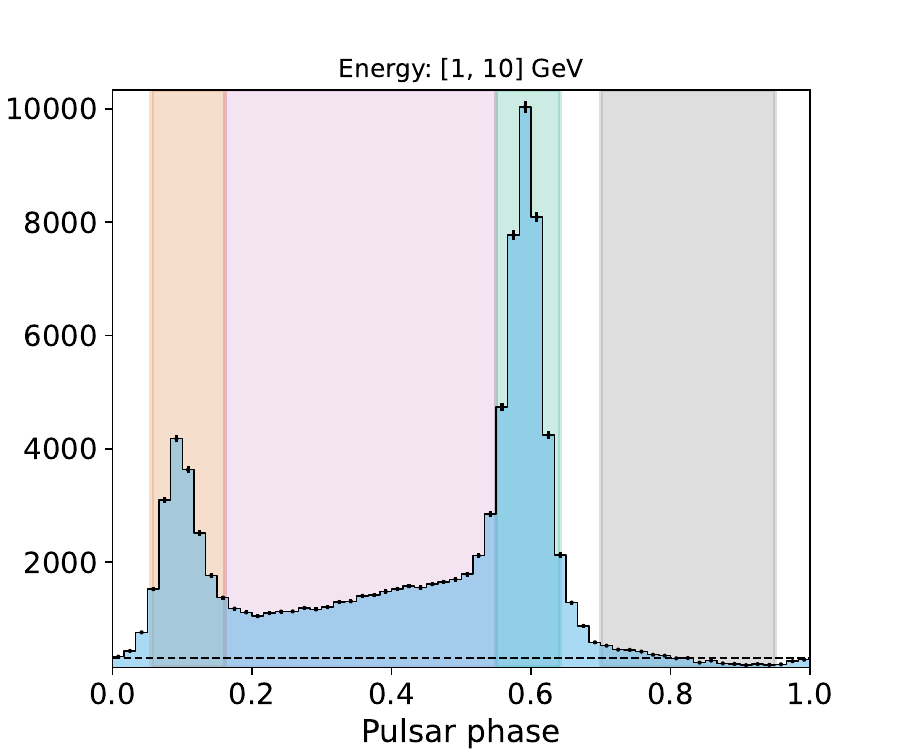}
    \includegraphics[width=0.4\linewidth]{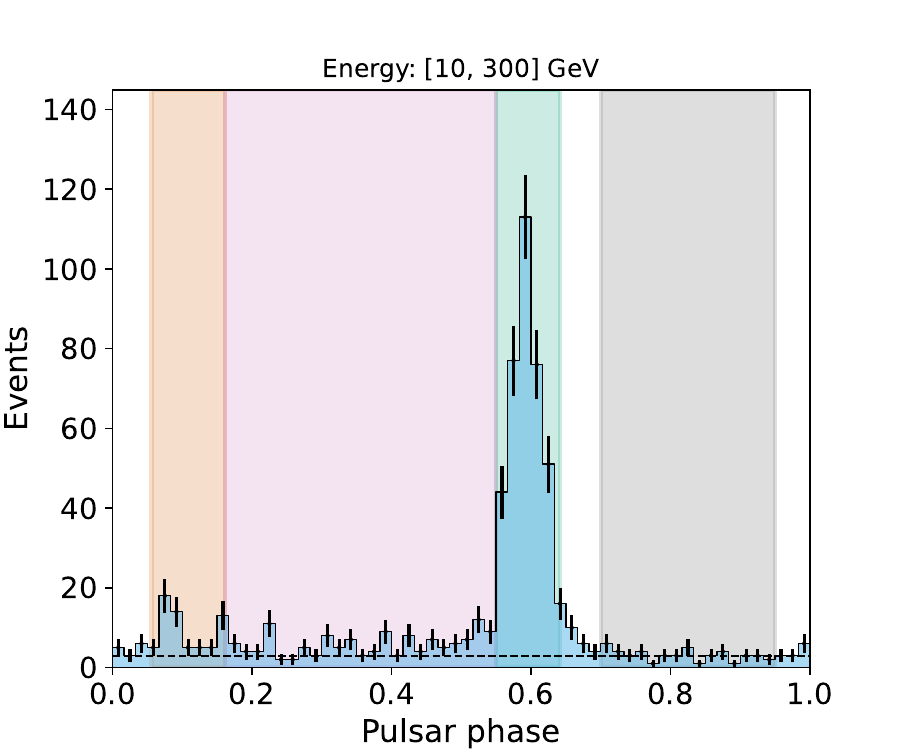}
    \includegraphics[width=0.4\linewidth]{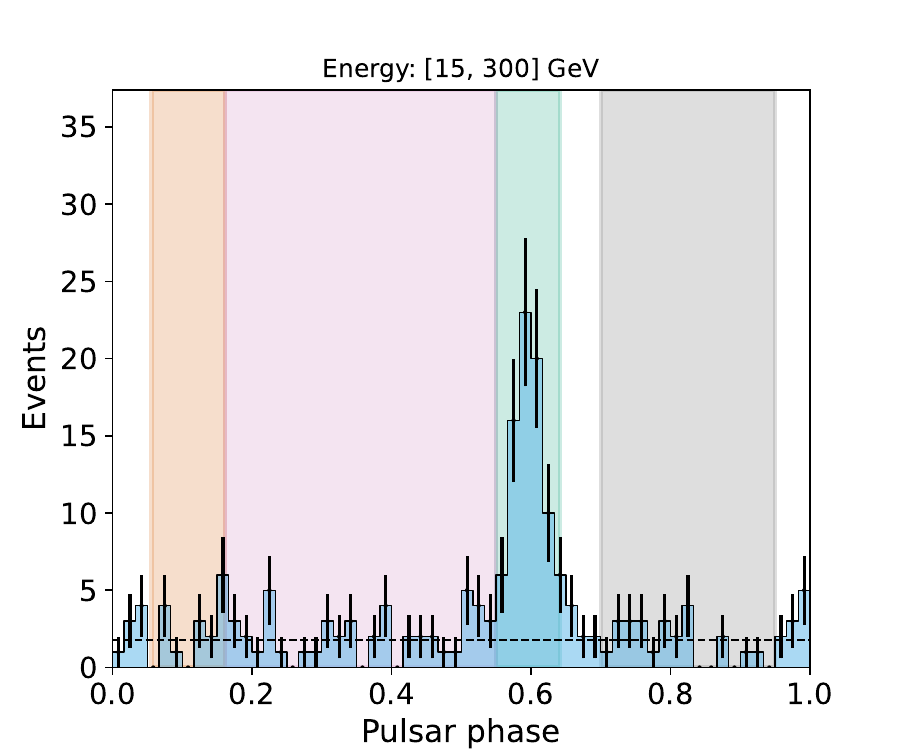}
    \caption{Phaseogram of Geminga obtained from the analysis of 16.6 years of \textit{Fermi}-LAT data using a circular extraction region with a 3$^\circ$ radius for four different energy ranges, indicated on top of each plot, and using the same colour code for the phase regions as Fig. \ref{fig:phaseogram}.}
    \label{fig:phaseogram_fermi}
\end{figure*}

The absence of emission from P1 above 15 GeV was already reported by \cite{Ackermann_2013} and later by \cite{Geminga_magic}. As a further comparison, we obtained the \textit{Fermi}-LAT phaseograms using a circular extraction region of 3$^\circ$ radius for different energy ranges, shown in Fig. \ref{fig:phaseogram_fermi}, to see how the two peaks evolve. The significance of both peaks and the Bridge is above 100$\sigma$ up to 10 GeV, but it decreases to 5.7$\sigma$, 23$\sigma$ and 4$\sigma$ above 10 GeV for P1, P2, and the Bridge, respectively. Above 15 GeV, only P2 is detected with a significance of 10.7$\sigma$, while P1 and the Bridge both show excesses at 1$\sigma$ level. The results obtained with the LST-1 are in line with these findings. However, it is possible to observe some hint of excess (significance above 1.5$\sigma$) from P1 and the Bridge regions in the LST-1 sample, as shown in Fig. \ref{fig:phaseogram}. These features will be investigated in the future when additional LSTs are operative.

\subsection{Morphology of P2}
We studied the shape of the second peak in the LST-1 phaseogram, fitting both a symmetric Gaussian and a symmetric Lorentzian model in phase. We chose symmetric models since no clear evidence of asymmetries is observed in P2 \citep{geminga_fermi}. We added a constant representing the background to each profile so that the final models could be written as
\begin{equation}
    f_{Gauss}(\phi)=A_G \exp \Bigg( - \frac{(\phi - \mu)^2}{2\sigma^2} \Bigg) + \mathrm{const_{Gauss}}
\end{equation}
for the Gaussian, and
\begin{equation}
    f_{Lor}(\phi)=A_L \ \Bigg[1+\bigg( \frac{\phi - x_0}{2\gamma} \bigg)^2 \Bigg]^{-1} + \mathrm{const_{Lor}}
\end{equation}
for the Lorentzian. In both cases, $\phi$ represents the value of the rotational phase.

We fitted these models in different reconstructed energy bins. We analysed the reconstructed energy range [15, 65] GeV, where the signal of P2 is concentrated. We further divided it into two logarithmically spaced bins, [15, 31] GeV and [31, 65] GeV, and repeated the analysis in both. We did not include data above 65 GeV due to the lack of signal from P2 at those energies. 

The best-fit results for the full band and the two energy bins are reported in Table \ref{tab:p2_fit_results} for both the Gaussian and the Lorentzian profiles, along with their statistical errors. All the statistical uncertainties from now on in the paper will be reported at 1$\sigma$ level. The full width at half maximum (FWHM) has been chosen as a measure of the peak width and it has been computed as $\textrm{FWHM}_{\rm Gauss} = 2\sigma \sqrt{2\log2}$ for the Gaussian and as $\textrm{FWHM}_{\rm Lor}=2\gamma$ for the Lorentzian. Henceforth, the notation `log' refers to the natural logarithm.

We found, for both models, that the position of the second peak of Geminga does not significantly change in the two studied energy bins. The value is compatible with that reported in \cite{phd_giovanni} within the statistical errors, but slightly differs from the best-fit values of \cite{geminga_fermi}. The same conclusions can be drawn for the peak width results.

The results obtained for the Gaussian and Lorentzian models are consistent with each other. To assess the goodness of fit of both models, we reported the associated p-value. We also computed the Akaike information criterion \citep{aic_criterion} defined as $\rm{AIC} = 2k - 2\log(\mathcal{L})$, where $k$ and $\mathcal{L}$ represent the number of model parameters and the maximum likelihood of the model. The $\rm AIC$ can be used to compare the results of two non-nested models and determine if one of the two is preferred. In general, lower coefficient values indicate a better agreement between the fit and the data. For the LST-1 sample, the Gaussian profile shows slightly lower values of the AIC. We computed the difference $\Delta \rm (AIC)$ for both the full band and the energy bins, obtaining $\Delta \rm (AIC)$ = (9.6, 2.2, 9.2) for the broad-band, the first and second bin, respectively.

As an additional cross-check, we fitted the phaseograms obtained with the \textit{Fermi}-LAT sample described in Sect. \ref{sec:fermi} for energies above 15 GeV (considered as the `broad-band' \textit{Fermi} sample) and in the same energy bins as for the LST-1 analysis, i.e. [15, 31] GeV and [31, 65] GeV. The results obtained, shown in Table \ref{tab:p2_fit_results}, are consistent with the LST-1 ones within the statistical uncertainties for both the Gaussian and the Lorentzian profiles. The fit in the second energy bin did not converge well, probably due to the low statistics of the \textit{Fermi}-LAT sample above 30 GeV, and for this reason we did not report the results. Also in this case, there is no significant preference between Gaussian and Lorentzian profiles, since both the p-values and the AIC in both the broad-band and the [15, 31] GeV fits are similar.

\begin{table*}[]
    \caption{Best-fit results for the P2 mean position and width (FWHM) for the symmetric Gaussian and Lorentzian profiles.}
    \label{tab:p2_fit_results}
    \resizebox{\textwidth}{!}{%
    \begin{tabular}{lccccccccc}
        \toprule
        \toprule
        Instrument & \multicolumn{1}{l}{} & \multicolumn{4}{c}{Gaussian} & \multicolumn{4}{c}{Lorentzian} \\
        \midrule
         & Energy (GeV) & $\mu$ & FWHM & p-value & AIC & $x_0$ & FWHM & p-value & AIC \\
        \midrule
        \multirow{3}{*}{CTAO LST-1} & {[15, 31]} & 0.595 $\pm$ 0.002 & 0.051 $\pm$ 0.004 & 0.91 & 51.3 & 0.597 $\pm$ 0.002 & 0.043 $\pm$ 0.005 & 0.62 & 60.9 \\
         & {[31, 65]} & 0.597 $\pm$ 0.004 & 0.05 $\pm$ 0.01 & 0.74 & 57.0 & 0.597 $\pm$ 0.004 & 0.04 $\pm$ 0.01 & 0.66 & 59.1 \\
         & Full band & 0.596 $\pm$ 0.002 & 0.050 $\pm$ 0.005 & 0.60 & 61.3 & 0.597 $\pm$ 0.002 & 0.041 $\pm$ 0.006 & 0.28 & 70.5 \\
        \midrule
        \multirow{2}{*}{\textit{Fermi}-LAT} & {[15, 31]} & 0.594 $\pm$ 0.001 & 0.053 $\pm$ 0.003 & 0.74 & 197.9 & 0.594 $\pm$ 0.001 & 0.050 $\pm$ 0.004 & 0.36 & 201.7 \\
         & >15 & 0.595 $\pm$ 0.001 & 0.047 $\pm$ 0.003 & 0.67 & 224.5 & 0.595 $\pm$ 0.001 & 0.045 $\pm$ 0.008 & 0.69 & 222.7 \\
        \midrule
        \begin{tabular}[c]{@{}l@{}}MAGIC\\ \small{(Ceribella, 2021)}
        \end{tabular} & {[25, 100]} & 0.597 $\pm$ 0.005 & 0.061 $\pm$ 0.012 & - & - & - & - & - & - \\
        \bottomrule
    \end{tabular}
    }
    \tablefoot{We show the results for both LST-1 and \textit{Fermi}-LAT, in the two energy bins and the full band, with the associated statistical uncertainties and the correspondent $AIC$ and p-value. As a comparison, the MAGIC results from \cite{phd_giovanni} are also included in the table. The results for the [31, 65] GeV fit of the \textit{Fermi}-LAT sample are not reported since the fit did not converge well.}
\end{table*}

\subsection{Spectral energy distribution of P2} \label{sec:sed}
In order to obtain the SED of the second peak, we performed a forward folding fit using \texttt{Gammapy} in the energy range [20, 95] GeV, considering five bins per decade, using a power law (PL) model:
\begin{equation}
    \frac{dN}{dE}=f_0 \ \Bigg( \frac{E}{E_0} \Bigg)^{-\Gamma},
\end{equation}
where $f_0$ represents the flux normalisation, $\Gamma$ is the spectral index, and $E_0$ is the reference energy.

The choice for the lower edge of the fitting range is connected to the estimated energy threshold of the analysis. We chose the upper edge of our energy binning closest to 100 GeV to assess the possibility of obtaining a spectral point at energies above 65 GeV. For more details, refer to Sect. \ref{sec:en_th}. We set the reference energy to $E_0=14.3$ GeV, the decorrelation energy computed with \texttt{Gammapy}, to have the lowest correlation between the parameters as well as the lowest statistical error on the flux normalisation.

The fit results with the power law model on P2 are reported in Table \ref{tab:pl_fit}, together with their statistical uncertainties. The spectrum observed by the LST-1 is slightly harder that of \cite{Geminga_magic}, but the two spectral indices are compatible if the systematic uncertainties are considered (see Sect. \ref{sec:syst}).

The flux points were calculated assuming the power law model fitted to the data, fixing the index and leaving free the normalisation in each bin, a procedure similar to the one explained in Sect. \ref{sec:fermi}. Both the LST-1 best-fit model and the flux points are depicted in Fig. \ref{fig:jointfit}, and the latter are also reported in Table \ref{tab:flux_points}. The signal from P2 is detected up to $\sim 65$ GeV, in the last energy bin we computed an upper limit. This is compatible with what has been found when studying the phaseogram (see Sect. \ref{sec:phaseogram}). 

\begin{table}[h!]
    \caption{Best-fit results of the spectral fitting of P2 with a power law model.}
    \label{tab:pl_fit}
    \centering
    \resizebox{0.85\hsize}{!}{%
    \begin{tabular}{cc}
        \toprule
        \toprule
        Parameter & Value \\
        \midrule
        $\Gamma$ & $4.5 \pm 0.4$\\
        $f_0$ &  $(9.99 \pm 0.75) \cdot 10^{-8}$ cm$^{-2}$ s$^{-1}$ TeV$^{-1}$\\
        \bottomrule
    \end{tabular}%
    }
    \tablefoot{Each parameter is shown with its statistical uncertainty. The reference energy is fixed to $E_0= 14.3$ GeV.}
\end{table}

\begin{table}[ht!]
    \caption{Flux points of the P2 SED obtained assuming the best-fit PL model of Tab. \ref{tab:pl_fit}.}
    \label{tab:flux_points}
    \resizebox{\hsize}{!}{%
    \begin{tabular}{cccc}
        \toprule
        \toprule
        E$_{ref}$ (GeV) & E$_{min}$ (GeV) & E$_{max}$ (GeV) & Flux (TeV $\cdot$ cm$^{-2}$ $\cdot$ s$^{-1}$) \\
        \midrule
        20 & 17 & 24 & (9.2 $\pm$ 1.1) $\cdot \ 10^{-12}$ \\
        28 & 24 & 33 & (3.5 $\pm$ 0.4) $\cdot \ 10^{-12}$ \\
        40 & 33 & 47 & (1.6 $\pm$ 0.4) $\cdot \ 10^{-12}$ \\
        56 & 47 & 67 & (1.2 $\pm$ 0.4) $\cdot \ 10^{-12}$ \\
        \bottomrule
    \end{tabular}%
    }
    \tablefoot{The upper limit on the last energy bin is not shown.}
\end{table}

\subsubsection{Energy threshold of the analysis and spectral binning} \label{sec:en_th}
To evaluate the analysis threshold we fitted with a Landau profile the energy distribution of the MCs employed in the data reduction process, in order to extract the position of the peak energy. The MCs are simulated with PL spectrum with index $\Gamma=2$, so we weighted the distribution for the spectral index of Geminga ($\Gamma = 4.5$). The energy threshold for the lowest-zenith MC determines the energy threshold of the analysis and it was found to be $\sim$ 10 GeV in true energy. This value is lower than the one reported in \cite{crab_lst} for the Crab pulsar study, due to the softness of the Geminga spectrum when compared to the Crab, and comparable with the performance of MAGIC's Sum-Trigger-II shown in \cite{Geminga_magic}. However, due to Geminga's spectrum being very steep, the bias, i.e. the mean difference between the reconstructed and the true energy, is considerably high around the threshold. Indeed, when producing the one-dimensional histogram of the MC distribution as a function of the reconstructed energy instead of the true energy, we found the peak was located around 20 GeV. This implies the energy threshold of 10 GeV in true energy corresponds to $\sim$ 20 GeV in reconstructed energy.

\begin{figure*}[!ht!]
    \centering
    \includegraphics[width=0.45\linewidth]{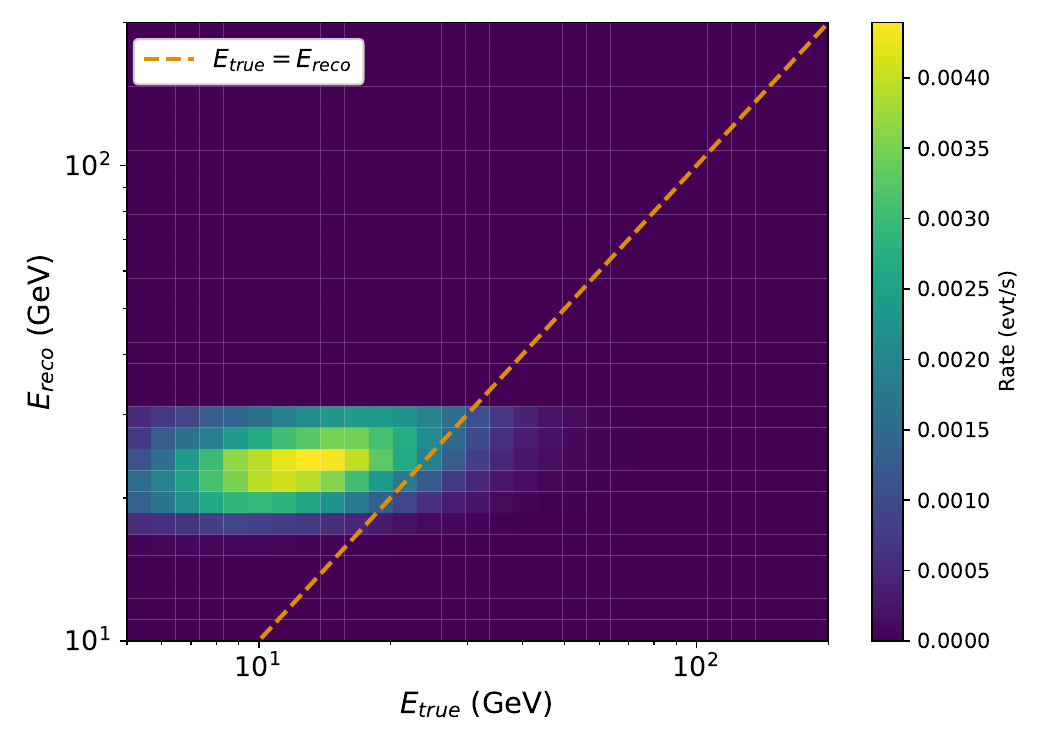}
    \includegraphics[width=0.45\linewidth]{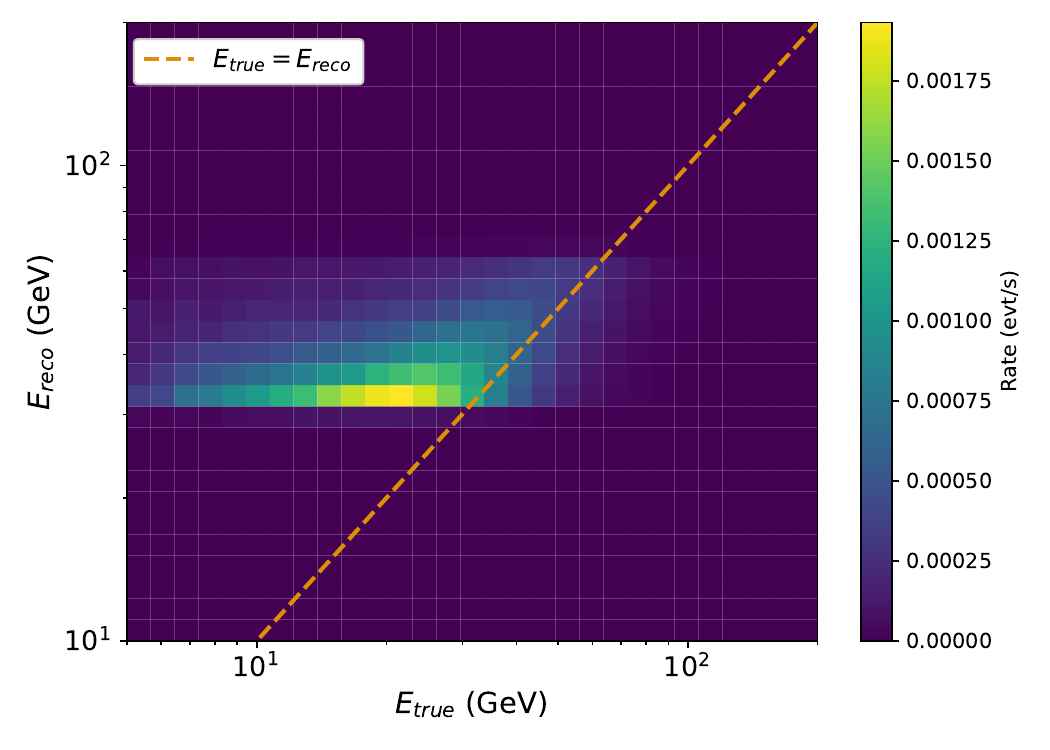}
    \caption{Two-dimensional weighted (considering Geminga's spectral index $\Gamma=4.5$) histogram of the reconstructed energy versus the true energy, projected onto the true energy axis for the two reconstructed energy bins used for the morphological study of P2, [15, 31] GeV (left) and [31, 65] GeV (right). The Monte Carlo data used for the plots were produced at $Zd=10^\circ$. The dashed line represents the equivalence between the true energy $E_{true}$ and the reconstructed energy $E_{reco}$. The z-axis is in units of rate, i.e. events per second.}
    \label{fig:migr_matr}
\end{figure*}

To better visualise the results, we produced the two-dimensional weighted histogram of the reconstructed energy versus the true energy, projected onto the true energy axis. This is the same as producing the one-dimensional weighted histogram of the MCs. The 2-D histogram, nevertheless, allows us to understand from which true energy the reconstructed events come from. We normalised the histogram to have a rate on the z-axis and plotted it for the two reconstructed energy bins used in the morphological study, i.e. [15, 31] and [31, 65] GeV. Figure \ref{fig:migr_matr} depicts the results. The Monte Carlo data chosen for Fig. \ref{fig:migr_matr} were simulated at $Zd=10^\circ$, the lowest zenith for the MC production, in order to obtain the plots for the lowest energy threshold possible.

For the energy binning of the \texttt{SpectrumDataset Gammapy} object representing the spectrum, we defined the true energy axis $E_{true}$ with 100 logarithmically spaced bins between 3 GeV and 50 TeV and the reconstructed energy axis $E_{reco}$ with 40 logarithmically spaced bins between 10 GeV and 10 TeV. The range for the spectral fit was defined in $E_{reco}$, due to the requirements of \texttt{Gammapy}, and it was chosen as a subset of the correspondent axis. We chose to start the subset from the edge closest to 20 GeV and end it at the edge closest to 100 GeV. This roughly corresponds to $\sim$ 10 - 100 GeV in true energy. For the flux point calculation, we set the first edge of the energy binning at $E_{reco} \simeq 17$ GeV to ensure the first flux point is around 20 GeV.

\subsection{Joint {\it Fermi}-LAT and LST-1 SED of P2}
\begin{figure}[!h!]
    \resizebox{\hsize}{!}{\includegraphics{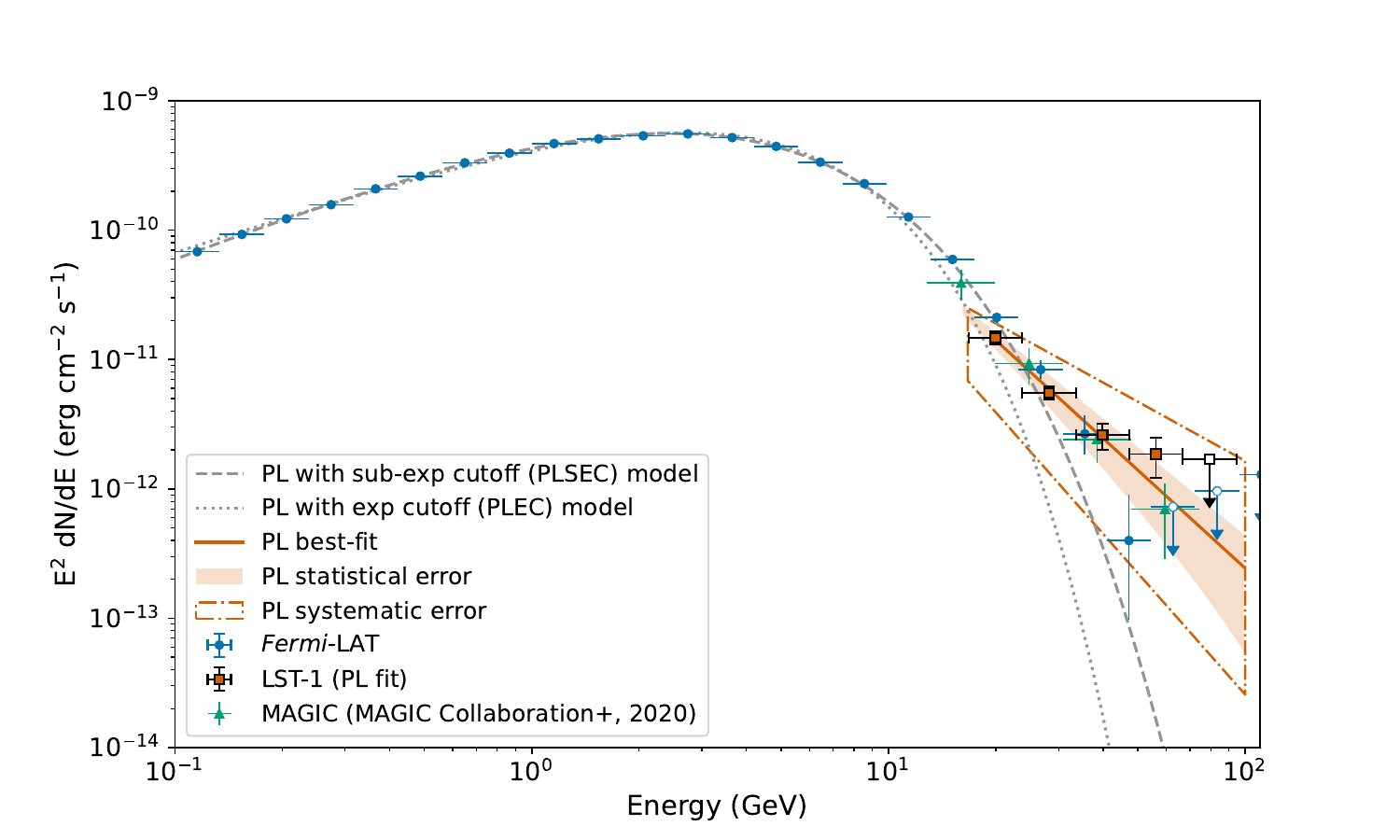}}
    \caption{Joint LST-1 (squares) and \textit{Fermi}-LAT (circles) data samples of P2, along with the best-fit results of both the power law with an exponential cut-off (PLEC, dotted line) and the power law with sub-exponential cut-off (PLSEC, dashed line). The power law fit of the LST-1 only points (orange squares) is shown together with its statistical 1$\sigma$ uncertainty band (solid line and shaded area) and the systematics uncertainty band (dash-dotted area), considering both the systematics on the index and the flux normalisation. The \cite{Geminga_magic} points are depicted as triangles for comparison. The horizontal error bars represent the width of the energy bins.}
    \label{fig:jointfit}
\end{figure}

To characterise the gamma-ray emission in a larger energy range we included in the fit the \textit{Fermi}-LAT flux points computed with the sample described in Sect. \ref{sec:fermi}. This is needed, in particular, to determine if a spectral cut-off exists and to verify if the emission at different energies is generated via the same mechanisms. 

We performed a joint fit of LST-1 and \textit{Fermi}-LAT data between 100 MeV and 100 GeV testing two spectral models. The first was a power law with an exponential cut-off (PLEC), and the second was a power law with a sub-exponential cut-off (PLSEC). The same mathematical law describes both models:
\begin{equation}
    \frac{dN}{dE}=f_0 \  \Bigg( \frac{E}{E_0} \Bigg)^{-\Gamma} \exp \Bigg[ - \Bigg( \frac{E}{E_c} \Bigg)^{\beta} \Bigg] = f_0 \  \Bigg( \frac{E}{E_0} \Bigg)^{-\Gamma} \exp \big[ - \big( \lambda E \big)^{\beta} \big],
\end{equation}
where $\beta=1$ in the case of PLEC, while $\beta<1$ for the PLSEC. $E_0$ is the reference energy and $E_c$ is the cut-off energy. We also report the formulation of the equation used in \texttt{Gammapy}, which uses the reciprocal of the cut-off energy $\lambda=1/E_c$.

The results of the joint fit are described in Table \ref{tab:jointfit} and the plot is shown in Fig. \ref{fig:jointfit}, along with the \cite{Geminga_magic} points for comparison. The value of the $\beta$ coefficient for the PLSEC model is compatible with that obtained by \cite{Geminga_magic} considering the statistical uncertainties.

\begin{table*}[ht]
    \centering
    \caption{Best-fit results for the joint fit of \textit{Fermi}-LAT and LST-1 data, for both tested models.}
    \label{tab:jointfit}
    \resizebox{0.9\hsize}{!}{%
    \begin{tabular}{cccccccc}
        \toprule
        \toprule
        Model & $f_0$ (TeV$^{-1}$ cm$^{-2}$ s$^{-1}$) & $E_0$ (GeV) & $\Gamma$ & $\lambda$ (GeV$^{-1}$) & $E_c$ (GeV) & $\beta$ & -$2\log \mathcal{L}$\\
        \midrule
        PLEC & (3.610 $\pm$ 0.002) $\cdot \ 10^{-4}$ & 1.00 & 1.079 $\pm$ 0.005 & 0.346 $\pm$ 0.003 & 2.890 $\pm$ 0.003 & 1 & 430.3 \\
        PLSEC & (6.1 $\pm$ 0.3) $\cdot \ 10^{-4}$ & 1.00 & 0.85 $\pm$ 0.02 & 0.76 $\pm$ 0.05 & 1.32 $\pm$ 0.05 & 0.73 $\pm$ 0.02 & 107.3 \\
        \bottomrule
    \end{tabular}%
    }
    \tablefoot{The critical energy $E_c$ is computed from its reciprocal $\lambda$ estimated with \texttt{Gammapy}.}
\end{table*}

Since the PLSEC and PLEC are nested models, Wilks theorem \citep{wilks} allows us to use of the likelihood ratio test to determine which is preferred. The PLEC model was found to be disfavoured with $-2 \Delta \log \mathcal{L}=323$. Considering a chi-square distribution with one degree of freedom, it corresponds to a rejection level of $TS\simeq 18\sigma$. This is in agreement with the findings already presented in the second pulsar catalogue (2PC) of \textit{Fermi}-LAT \citep{2pc_fermi} and later confirmed in the 3PC \citep{Smith_2023}, which found a preference in pulsar spectra for a softer sub-exponential cut-off rather than the classical exponential. This is also compatible with what has been found by \cite{Geminga_magic}. However, it is clear from the plot that the PLSEC still does not match well with the LST-1 flux points, even though it is preferred over the PLEC. To better investigate the preference for a non-curved model over a curved one, we performed a test to assess the presence of curvature (see Sect. \ref{sec:curvature}).

\subsection{Testing for curvature in the spectrum} \label{sec:curvature}
We used the \textit{Fermi}-LAT flux points to verify the presence of a possible curvature in the spectrum at energies above 10 GeV. To do so, we fitted the LST-1 and \textit{Fermi} points using a power law and a log parabola, the latter described by the following parametrisation: 
\begin{equation}
    \frac{dN}{dE} = f_0 \Bigg( \frac{E}{E_0} \Bigg)^{-a -b \log(E/E_0)},
\end{equation}
where the parameter $b$ represents the curvature index, and it is equal to zero for a non-curved model (i.e. a simple power law). We fitted both models to the joint dataset imposing a minimum energy of the \textit{Fermi}-LAT points of 10, 15 and 20 GeV. We also performed this test on the LST-1-only sample, using 20 and 25 GeV as minimum energies for the fit.

When considering the joint \textit{Fermi}-LAT and LST-1 sample, a log parabola with concave (negative) curvature (i.e. positive $b$) is strongly favoured over the power law when setting a minimum energy of 10 GeV and 15 GeV, with $\Delta TS =10\sigma$ for the former and $\Delta TS =8.9\sigma$ for the latter. In both cases, the best-fit curvature parameter is $b=3.3 \pm 0.6$. The preference for a curved log parabola model decreases to a 4.6$\sigma$ level for $E_{min}=20$ GeV, for which $b=1.95 \pm 0.99$ is derived. We also tested $E_{min}=25$ GeV and obtained again a slight preference for the log parabola at 2.6$\sigma$ level, with best-fit curvature parameter $b=0.9 \pm 0.3$, almost consistent with zero. The overall preference for a curved log parabola model is probably related to the larger number of \textit{Fermi}-LAT points when compared to those of LST-1, especially in the range 10 - 20 GeV that is not covered by the latter instrument. This preference tends to decrease as $E_{min}$ increases.

For the \textit{Fermi}-LAT only sample, instead, we obtained a 3.5$\sigma$ preference for the log parabola and a best fit $b=1.0 \pm 0.3$ when setting $E_{min}=10$ GeV, and only a 1$\sigma$ preference for a log parabola with $b$ always consistent with zero in the case of $E_{min}=15, \ 20$ GeV. No curvature seems to be detected at higher energies and it is likely due to the low statistics of the sample in the correspondent energy range.

Similarly it happens in the case of the LST-1 only sample, for which the results obtained fitting a log parabola for both minimum energies return a value of $b$ which is consistent with zero and a $-2 \Delta \log \mathcal{L} \ll 1$ when compared to the power law. This implies no statistical preference between the two models and no curvature is detected in the LST-1 sample. However, a non-detection does not necessarily mean the curvature is absent for a wider energy range. With the current LST-1 sample, we cannot discard either possibility, but it may be possible to understand more about it by collecting more data to lower the statistical uncertainty and extending it to higher energies.

\section{Evaluation of the systematic effects} \label{sec:syst}
The best-fit result for the PL fitting of the P2 spectrum we found was slightly harder than the spectrum reported by \cite{Geminga_magic}. To assess the compatibility of MAGIC's results with ours, we examined the contribution of the systematic effects in the analysis of the LST-1-only sample. Different factors can contribute to the systematic uncertainties, with a more or less significant impact on the final result, and we investigated the most relevant ones. This is the first time such a detailed study of the systematics is performed for LST-1 analysis.

\subsection{\textit{Different MC efficiency}}
The first factor we studied was the setting of different energy-dependent selection efficiencies of the MC for both alpha and gammaness. The value of 70\% is usually chosen as the standard because it is a good compromise between having a large number of events classified as gamma-originated and a reliable gamma-hadron separation. To test whether and how the choice of efficiency impacts the final results, we computed the spectral parameters in a grid of alpha efficiencies of 70\% and 90\% and gammaness efficiencies of 40\%, 70\%, and 90\%. 

We found that both the spectral index and the flux normalisation change within 5\% of the reference value, with a maximum positive absolute variation of $\Gamma$ of +0.13 and a maximum negative absolute variation of -0.02. We can conclude that this effect is subdominant to the statistical uncertainties.

\subsection{\textit{Mismatches in the reconstruction}}
Another important systematic effect that must be taken into account is related to the mismatches between the MC and the observed data, which lead to a wrong reconstruction of the events. One of the most common ones is the energy scale mismatch considered here. Non-ideal conditions during the observations, such as dirty mirrors or lower atmospheric transparency, affect the total amount of recorded light by the system, introducing a bias in the reconstruction, in particular, of the energy. The proper way to simulate this effect would be to have dedicated MC productions in which an energy bias is introduced. Still, this method is time-consuming and requires significant resources. Defining the energy scaling as $\varepsilon$, we can say that, for small values of $\varepsilon$, the effect of the energy bias can be approximately reproduced by shifting the migration matrix along the true energy axis. 

We considered $\varepsilon = \pm$15\%, which is the typical value used for the studies of energy scale systematics for IACTs \citep{performance_magic}. We produced the scaled IRFs in the two cases and derived the spectrum again. We found that the spectral index is not significantly affected by the energy scaling, while the flux normalisation changes even up to $\sim$ 55\% as a consequence of the shift of the spectrum in the energy axis. In detail, the absolute positive variation of $\Gamma$ is +0.04 and the negative is -0.08. The overall effect is a rigid shift of the PL to the right or the left depending on the scaling.

The same type of systematic uncertainty has been considered in \cite{Geminga_magic}, for which they reported a $\sim$1\% contribution for the spectral index, corresponding to an absolute variation of $\sim$ 0.06. If we only consider this effect, the results of LST-1 are still not consistent with MAGIC's.

\subsection{\textit{Different Zd cut of the sample}}
The choice of the zenith angle cut of the data impacts the overall energies of the events: the larger the applied zenith cut, the higher the energy threshold of the data sample will be. We performed the analysis using a zenith cut of $Zd<50$° to have the largest possible sample, but $\sim90$\% of the observation time ($\sim$ 54 hours out of 60) data were taken at $Zd<25$°, assuring the lowest possible energy threshold for the sample. In Table \ref{tab:zeniths}, the observational hours for each of the $Zd$ cuts are shown.

\begin{table}[h!]
    \caption{Total observation time of the LST-1 sample for the different maximum zenith distance ($Zd$) cuts applied for the analysis of the systematic effects.}
    \resizebox{\hsize}{!}{%
        \begin{tabular}{cc|cc}
        \toprule
        \toprule
        Max \textit{Zd} (deg) & Total time (h) & Max \textit{Zd} (deg) & Total time (h) \\ \midrule
        50 & 60.1 & 25 & 54.1 \\
        40 & 59.5 & 20 & 42.8 \\
        35 & 58.7 & 15 & 26.8  \\
        30 & 56.6 & &  \\
        \bottomrule
    \end{tabular}%
    }
    \label{tab:zeniths}
\end{table}

We obtained the spectrum for different values of the zenith cut of the sample and estimated how much the choice of $Zd$ influences $\Gamma$ and $f_0$. The most significant variation is obtained when setting $Zd<20$°. In this case, we obtain an absolute negative variation in the index of -0.5 and a relative positive variation in the normalisation of +12\%. The other variations obtained with this test are much smaller.

\subsection{\textit{Different intensity cut of the sample}}
The last investigated systematic effect is related to the choice of the intensity cut applied to both simulated and observed data. This cut is usually applied to guarantee an overall good match between simulations and observations, and the standard value adopted for LST-1 analysis is an image intensity above 50 p.e. as shown in \cite{performance_paper}. We tried a series of different cut values, from a minimum intensity of 20 p.e. to a maximum of 70 p.e., and computed the spectral results in each case.

In this case, the positive and negative absolute variations of the spectral index are +0.2 and -0.3, respectively. The uncertainties are around 20\% for the normalisation.

\subsection{\textit{Total contribution}}
To compute the total contribution of the systematic uncertainties on the spectral index, we assumed the correlation between all the examined effects to be low. In this way, we could sum in quadrature all the maximum positive and negative variations to obtain the total positive and negative uncertainty. We derived a final result for the spectral index of 
\begin{center}
    $\Gamma= (4.5 \pm 0.4_{stat})^{+0.2_{sys}}_{-0.6_{sys}}$.
\end{center}

This result is compatible with the one reported by \cite{Geminga_magic}, considering the statistical uncertainty of the MAGIC result. This is the first time another facility has cross-checked the MAGIC Collaboration's result and it confirms again the potential of the LSTs.

The systematic uncertainties on the normalisation are much larger, $\Delta f_{0, sys}^+=+ 6 \cdot 10^{-8}$ cm$^{-2}$ s$^{-1}$ TeV$^{-1}$ and $\Delta f_{0, sys}^-=-5 \cdot 10^{-8}$ cm$^{-2}$ s$^{-1}$ TeV$^{-1}$. This is expected since the spectrum of Geminga is very steep and small fluctuations of the spectral index can induce large variations of the normalisation. The total systematic uncertainty band, considering both the uncertainty on the index and normalisation, is depicted in Fig. \ref{fig:jointfit} and in Fig.\ref{fig:harding_comparison} as the dash-dotted area.

\section{Discussion and conclusions} \label{sec:final}
In this paper, we report the detection of the Geminga pulsar by the LST-1 of CTAO. The signal from the second peak of the phaseogram, P2, is detected with a significance of 12$\sigma$, achieved with 60 hours of good-quality data. The first peak, P1, remains undetected, even though a $2.6\sigma$ excess has been observed. We also observed an excess of $1.8\sigma$ for the first inter-peak (Bridge) region.

The detection of P1 with the LST-1 alone is still challenging. Considering the evolution of the significance with time we derived in Sect. \ref{sec:phaseogram}, more than 200 hours would be needed to achieve a 5$\sigma$ detection. However, with the deployment of the full array of four LSTs at CTAO-North, it will be possible to reduce this time and detect P1, and possibly the Bridge, in the near future. Using the latest \texttt{Prod5} IRFs of CTAO \citep{ctao_irfs} and an extrapolated power law spectrum from the \textit{Fermi}-LAT data above 5 GeV, we estimated around 30 hours will be needed for a 5$\sigma$ detection of the first peak with the full array of LSTs.

We performed a morphological study of the second peak by testing two models, the symmetric Gaussian and the symmetric Lorentzian. We found compatible peak mean positions and FWHMs between the two profiles and no energy-dependent evolution in the two logarithmically spaced energy bins considered, [15, 31] GeV and [31, 65] GeV. The Gaussian model, in general, returns better goodness-of-fit values, highlighting a slight preference for this profile. More tests on the pulse morphology will be carried out after new observations, possibly extending the study to P1 and the Bridge, to further investigate the geometry of the peaks.

We then extracted the SED of P2 in the [20, 95] GeV range, obtaining four flux points and an upper limit in the last energy bin. As also observed with the study of the phaseogram, the signal from P2 is reconstructed up to 65 GeV. The best-fit result yielded a power law with spectral index $\Gamma=(4.5\pm0.4_{stat})^{+0.2_{sys}}_{-0.6_{sys}}$, compatible with the previous result of \cite{Geminga_magic}, and a normalisation at 14.3 GeV of $f_0=(9.99 \pm 0.75_{stat})^{+6_{sys}}_{-5_{sys}} \cdot 10^{-8}$ cm$^{-2}$ s$^{-1}$ TeV$^{-1}$. For the first time, a result on a pulsar other than the Crab is being cross-checked and confirmed by two different IACTs. This is important in the field of pulsars detected by IACTs, since only a few sources have been detected and it is not straightforward to perform cross-checks between different instruments. 

We also analysed 16.6 years of \textit{Fermi}-LAT data to perform a joint fit with the LST-1 sample and extend the Geminga spectrum down to 100 MeV. Two models, a power law with exponential or sub-exponential cut-off, were tested and we found that the exponential cut-off model can be rejected at an 18$\sigma$ level, in line with previous findings by \textit{Fermi}-LAT and MAGIC. However, even if the power law with sub-exponential cut-off returns a better fit, it does not correctly describe the spectrum obtained with LST-1.

For this reason, using the \textit{Fermi} flux points, we tested for the presence of curvature in the high-energy end of the spectrum. We performed a likelihood ratio test between a power law and a log parabola spectral model using different minimum energies for the \textit{Fermi}-LAT data, and we repeated the test on the LST-1-only sample. For the latter, no curvature is detected, probably due to the limited statistics and sampled energy range. When considering the joint dataset, the log parabola is found to be statistically favoured over the power law. However, the preference tends to decrease as the minimum energy set as the starting point of the fit increases. Nevertheless, the fit results may be affected by the absence of LST-1 flux points below 20 GeV or by the low statistics of both samples at the highest energies.

After the early measurements by \textit{Fermi}-LAT, theories based on the outer gap (OG) concept were favoured \citep{Cheng_1986, Romani_1996}. To explain the gamma-ray emission from pulsars, these theories assumed that particles were accelerated to relativistic velocities inside charge-depleted regions high up in the magnetosphere. However, in addition to the uncertainties and approximations used by these models (see \cite{Vigano_Torres_Hirotani_2015a} for a discussion), particle-in-cell (PIC) and force-free electrodynamics simulations signalled that instead of gaps, the most likely location of the acceleration of the gamma-ray-emitting particles is the so-called Y-point, a localised region at or beyond the light cylinder, see \cite{Cerutti_2019} and references therein for details. Also, particularly after the second \textit{Fermi}-LAT pulsar catalogue \citep{2pc_fermi}, it was clear that the variety of the \textit{Fermi}-LAT spectra, even without considering VHE tails, was already beyond what curvature-only models can describe. \cite{Vigano_Torres_Hirotani_2015b} proposed a synchro-curvature (SC) approach to tackle this this variety of spectra. This was used to provide fits of the spectra of gamma and X-ray pulsars \citep{Torres_2018}, and when coupled to light curves predictions, to describe their global properties as well \citep{Iniguez_2024}. \cite{Harding_2021, Kalapotharakos_2018, Brambilla_2018} and \cite{Cerutti_2025}, among a few others, introduced PIC-motivated models. In all these cases, though, the broad-band emission, particularly beyond tera-electronvolt energies, seems to require Compton components.

\cite{Harding_2021} proposed to describe the optical to X-ray emission with synchrotron radiation, SC (as used above) up to $\sim 100$ GeV and inverse Compton scattering (ICS) above tera-electronvolt energies. They also include a synchrotron-self-Compton (SSC) component, but, except for the Crab pulsar, the SC emission completely outshines it. The ICS component at tera-electronvolt energies, instead, could in principle be detected. The \cite{Harding_2021} model comprises several parameters, such as the inclination angle $\alpha$ between the rotational axis and the magnetic field, the viewing angle $\zeta$ between the pulsar and the observer, and the low and high values of the accelerated electric field, the latter expressed through the parameters R$^{low}_{acc}=e\text{E}_{\parallel}^{low}/mc^2$ and R$^{high}_{acc}=e\text{E}_{\parallel}^{high}/mc^2$. The values of these parameters are adjusted to match the observations, in particular ${E}_{\parallel}^{low}$ and ${E}_{\parallel}^{high}$ are chosen to reproduce the hard X-rays to giga-electronvolt data and $\zeta$ is fixed to the value that best matches the \textit{Fermi}-LAT and VHE data. After setting the values of those parameters, the ICS and the total SED are computed.

\begin{figure}[!h!]
    \centering
    \resizebox{\hsize}{!}{\includegraphics{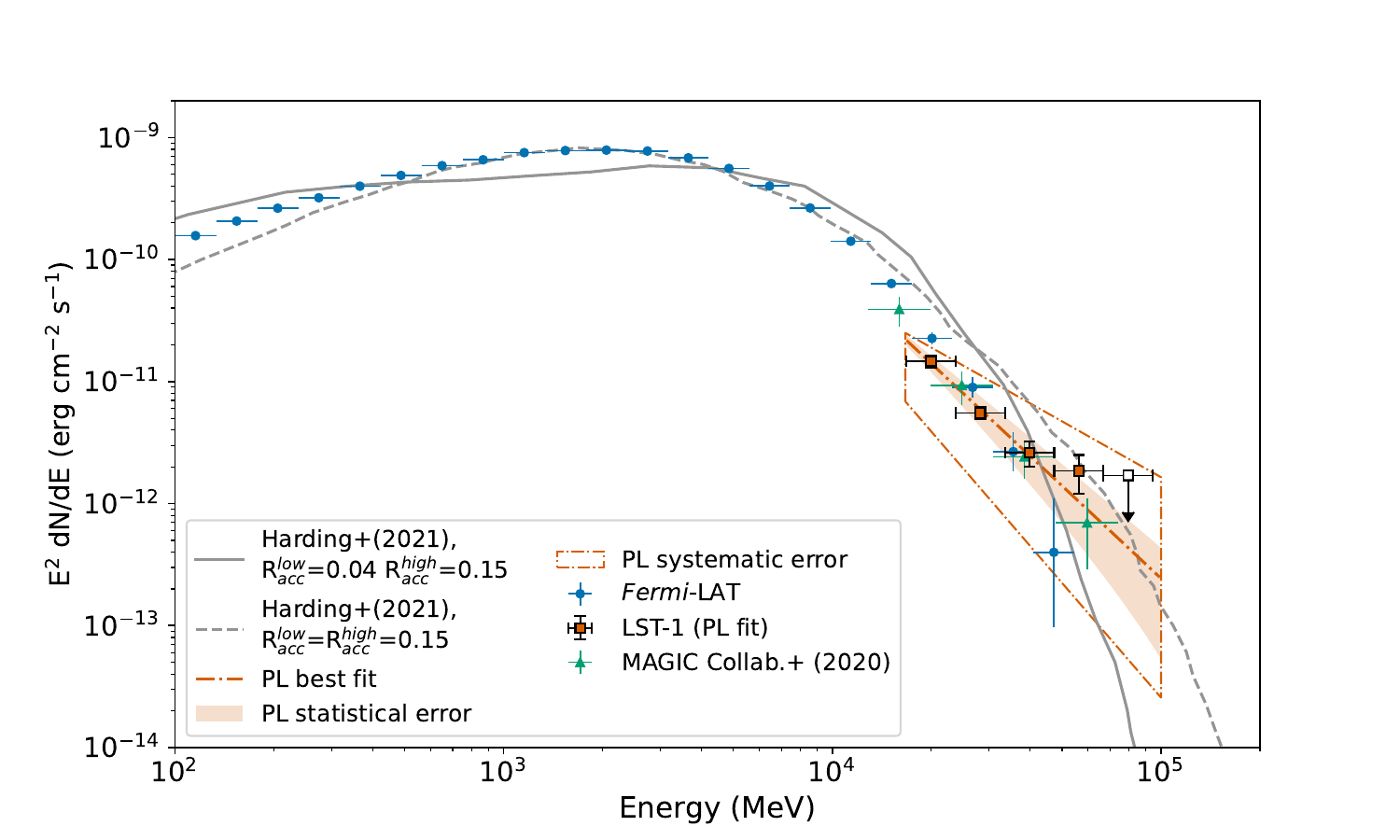}}
    \caption{Adaptation of Fig. 10 of \cite{Harding_2021} showing the two models derived in the paper for the SC emission of Geminga, together with the LST-1 P2 SED obtained in this work (same as Fig. \ref{fig:jointfit}). The \textit{Fermi}-LAT and MAGIC points from the analysis presented in this work and \cite{Geminga_magic}, respectively, are also reported for comparison. Note that the \textit{Fermi}-LAT points represent the phase-averaged flux, i.e. they take into account the emission from both P1 and P2.}
    \label{fig:harding_comparison}
\end{figure}

For the specific case of Geminga, the SED model was obtained for $\alpha=75^\circ$, $\zeta=55^\circ$ and for two different sets of values for the low and high accelerating electric field, the first one with R$^{low}_{acc}$ = 0.04 and R$^{high}_{acc}$ = 0.15 and the second with R$^{low}_{acc}$ = R$^{high}_{acc}$ = 0.15. As one can see from Fig. 10 of \cite{Harding_2021}, the dominant component of the high-energy emission of Geminga is that of SC radiation. Moreover, the predicted ICS is too faint to be detected, even considering CTAO-North's sensitivity. This is caused by the larger radius of the light cylinder due to the lower rotational speed of Geminga when compared to the younger pulsars (i.e. Vela, Crab) considered in \cite{Harding_2021}, which implies a larger distance between the current sheet and the neutron star producing the X-ray photon seeds for the ICS. In this context, the spectrum observed both by MAGIC and LST-1 is a continuous tail-like extension of the \textit{Fermi}-LAT one and is interpreted as curvature radiation emission, without the need for additional components. We depicted in Fig. \ref{fig:harding_comparison} the LST-1 SED together with the SC component of Geminga predicted by this model for the two cases of R$^{low}_{acc}$, R$^{high}_{acc}$ considered in the reference, as explained above. The model describes the emission of both peaks together, so we included in the same Figure the \textit{Fermi}-LAT flux points of the phase-averaged analysis. In the case of MAGIC and LST-1, instead, we considered the flux points of P2 only, since the emission above 15 GeV is fully dominated by the second peak. The \cite{Harding_2021} model also predicts the absence of signal from P1 above 20 GeV, which is compatible with what has been found in this work. All taken into account, this model could explain some phenomenology of the Geminga emission from the \textit{Fermi}-LAT to the LST-1 energy range, even though the match with the LST-1 SED and the higher energy flux points of \textit{Fermi}-LAT could be improved. With the present sample, it is not possible to completely rule out this or other theoretical interpretations and many open questions remain. Further studies, possibly with more LSTs, are necessary to validate or reject the model.

Overall, we can conclude that the results obtained on the Geminga pulsar prove the LST-1 to be an excellent telescope for pulsar observations at the upper end of their spectra, also considering its partial overlap with the \textit{Fermi}-LAT energy range. With this study, we were able to prove that the threshold of the telescope for sources with very steep spectra can be as low as 10 GeV, even though a large bias is observed at energies below 20 GeV. This issue will be reduced when at least another LST becomes operational because the use of a stereoscopic system improves the reconstruction of the events. With the full array of LSTs, the sensitivity will be improved and more tests could be performed, such as investigating the behaviour of P1 and of the bridge as a function of energy, as well as a better determination of the P2 spectrum to verify the validity of the theoretical models.

\begin{acknowledgements}
    We gratefully acknowledge financial support from the following agencies and organisations:

    Conselho Nacional de Desenvolvimento Cient\'{\i}fico e Tecnol\'{o}gico (CNPq), Funda\c{c}\~{a}o de Amparo \`{a} Pesquisa do Estado do Rio de Janeiro (FAPERJ), Funda\c{c}\~{a}o de Amparo \`{a} Pesquisa do Estado de S\~{a}o Paulo (FAPESP), Funda\c{c}\~{a}o de Apoio \`{a} Ci\^encia, Tecnologia e Inova\c{c}\~{a}o do Paran\'a - Funda\c{c}\~{a}o Arauc\'aria, Ministry of Science, Technology, Innovations and Communications (MCTIC), Brasil;
    Ministry of Education and Science, National RI Roadmap Project DO1-153/28.08.2018, Bulgaria;
    Croatian Science Foundation (HrZZ) Project IP-2022-10-4595, Rudjer Boskovic Institute, University of Osijek, University of Rijeka, University of Split, Faculty of Electrical Engineering, Mechanical Engineering and Naval Architecture, University of Zagreb, Faculty of Electrical Engineering and Computing, Croatia;
    Ministry of Education, Youth and Sports, MEYS  LM2023047, EU/MEYS CZ.02.1.01/0.0/0.0/16\_013/0001403, CZ.02.1.01/0.0/0.0/18\_046/0016007, CZ.02.1.01/0.0/0.0/16\_019/0000754, CZ.02.01.01/00/22\_008/0004632 and CZ.02.01.01/00/23\_015/0008197 Czech Republic;
    CNRS-IN2P3, the French Programme d’investissements d’avenir and the Enigmass Labex, 
    This work has been done thanks to the facilities offered by the Univ. Savoie Mont Blanc - CNRS/IN2P3 MUST computing center, France;
    Max Planck Society, German Bundesministerium f{\"u}r Bildung und Forschung (Verbundforschung / ErUM), Deutsche Forschungsgemeinschaft (SFBs 876 and 1491), Germany;
    Istituto Nazionale di Astrofisica (INAF), Istituto Nazionale di Fisica Nucleare (INFN), Italian Ministry for University and Research (MUR), and the financial support from the European Union -- Next Generation EU under the project IR0000012 - CTA+ (CUP C53C22000430006), announcement N.3264 on 28/12/2021: ``Rafforzamento e creazione di IR nell’ambito del Piano Nazionale di Ripresa e Resilienza (PNRR)'';
    ICRR, University of Tokyo, JSPS, MEXT, Japan;
    JST SPRING - JPMJSP2108;
    Narodowe Centrum Nauki, grant number 2019/34/E/ST9/00224, Poland;
    The Spanish groups acknowledge the Spanish Ministry of Science and Innovation and the Spanish Research State Agency (AEI) through the government budget lines
    PGE2022/28.06.000X.711.04,
    28.06.000X.411.01 and 28.06.000X.711.04 of PGE 2023, 2024 and 2025,
    and grants PID2019-104114RB-C31,  PID2019-107847RB-C44, PID2019-104114RB-C32, PID2019-105510GB-C31, PID2019-104114RB-C33, PID2019-107847RB-C43, PID2019-107847RB-C42, PID2019-107988GB-C22, PID2021-124581OB-I00, PID2021-125331NB-I00, PID2022-136828NB-C41, PID2022-137810NB-C22, PID2022-138172NB-C41, PID2022-138172NB-C42, PID2022-138172NB-C43, PID2022-139117NB-C41, PID2022-139117NB-C42, PID2022-139117NB-C43, PID2022-139117NB-C44, PID2022-136828NB-C42, PDC2023-145839-I00 funded by the Spanish MCIN/AEI/10.13039/501100011033 and “and by ERDF/EU and NextGenerationEU PRTR; the "Centro de Excelencia Severo Ochoa" program through grants no. CEX2019-000920-S, CEX2020-001007-S, CEX2021-001131-S; the "Unidad de Excelencia Mar\'ia de Maeztu" program through grants no. CEX2019-000918-M, CEX2020-001058-M; the "Ram\'on y Cajal" program through grants RYC2021-032991-I  funded by MICIN/AEI/10.13039/501100011033 and the European Union “NextGenerationEU”/PRTR and RYC2020-028639-I; the "Juan de la Cierva-Incorporaci\'on" program through grant no. IJC2019-040315-I and "Juan de la Cierva-formaci\'on"' through grant JDC2022-049705-I. They also acknowledge the "Atracci\'on de Talento" program of Comunidad de Madrid through grant no. 2019-T2/TIC-12900; the project "Tecnolog\'ias avanzadas para la exploraci\'on del universo y sus componentes" (PR47/21 TAU), funded by Comunidad de Madrid, by the Recovery, Transformation and Resilience Plan from the Spanish State, and by NextGenerationEU from the European Union through the Recovery and Resilience Facility; “MAD4SPACE: Desarrollo de tecnolog\'ias habilitadoras para estudios del espacio en la Comunidad de Madrid" (TEC-2024/TEC-182) project funded by Comunidad de Madrid; the La Caixa Banking Foundation, grant no. LCF/BQ/PI21/11830030; Junta de Andaluc\'ia under Plan Complementario de I+D+I (Ref. AST22\_0001) and Plan Andaluz de Investigaci\'on, Desarrollo e Innovaci\'on as research group FQM-322; Project ref. AST22\_00001\_9 with funding from NextGenerationEU funds; the “Ministerio de Ciencia, Innovaci\'on y Universidades”  and its “Plan de Recuperaci\'on, Transformaci\'on y Resiliencia”; “Consejer\'ia de Universidad, Investigaci\'on e Innovaci\'on” of the regional government of Andaluc\'ia and “Consejo Superior de Investigaciones Cient\'ificas”, Grant CNS2023-144504 funded by MICIU/AEI/10.13039/501100011033 and by the European Union NextGenerationEU/PRTR,  the European Union's Recovery and Resilience Facility-Next Generation, in the framework of the General Invitation of the Spanish Government’s public business entity Red.es to participate in talent attraction and retention programmes within Investment 4 of Component 19 of the Recovery, Transformation and Resilience Plan; Junta de Andaluc\'{\i}a under Plan Complementario de I+D+I (Ref. AST22\_00001), Plan Andaluz de Investigaci\'on, Desarrollo e Innovación (Ref. FQM-322). ``Programa Operativo de Crecimiento Inteligente" FEDER 2014-2020 (Ref.~ESFRI-2017-IAC-12), Ministerio de Ciencia e Innovaci\'on, 15\% co-financed by Consejer\'ia de Econom\'ia, Industria, Comercio y Conocimiento del Gobierno de Canarias; the "CERCA" program and the grants 2021SGR00426 and 2021SGR00679, all funded by the Generalitat de Catalunya; and the European Union's NextGenerationEU (PRTR-C17.I1). This research used the computing and storage resources provided by the Port d’Informaci\'o Cient\'ifica (PIC) data center.
    State Secretariat for Education, Research and Innovation (SERI) and Swiss National Science Foundation (SNSF), Switzerland;
    The research leading to these results has received funding from the European Union's Seventh Framework Programme (FP7/2007-2013) under grant agreements No~262053 and No~317446;
    This project is receiving funding from the European Union's Horizon 2020 research and innovation programs under agreement No~676134;
    ESCAPE - The European Science Cluster of Astronomy \& Particle Physics ESFRI Research Infrastructures has received funding from the European Union’s Horizon 2020 research and innovation programme under Grant Agreement no. 824064.\\

    \textit{Author contribution}:
    G. Brunelli:  project coordination, LST-1 pulsar analysis (phaseogram and SED), \textit{Fermi}-LAT phaseograms, joint fit of LST-1 and \textit{Fermi}-LAT, systematic error estimation, results discussion.
    A. Mas-Aguilar: analysis coordination, cross-check on LST-1 pulsar analysis (phaseogram and SED), cross-check on joint fit of LST-1 and \textit{Fermi}-LAT, systematic error estimation, results discussion.
    G. Ceribella: \textit{Fermi}-LAT analysis, ephemeris production, results discussion.
    P. K. H. Yeung: cross-check on LST-1 pulsar analysis (phaseogram and SED), cross-check of the \textit{Fermi}-LAT phaseogram, results discussion.
    R. Lopez-Coto: PI of the observations, project and analysis coordination, results discussion.
    M. Lopez-Moya: \textit{Fermi}-LAT analysis, results discussion.
    All corresponding authors have participated in the paper drafting and edition. The rest of the authors have contributed in one or several of the following ways: design, construction, maintenance and operation of the instrument(s) used to acquire the data; preparation and/or evaluation of the observation proposals; data acquisition, processing, calibration and/or reduction; production of analysis tools and/or related Monte Carlo simulations; discussion and approval of the contents of the draft.
\end{acknowledgements}

\bibliographystyle{aa} 
\bibliography{Bibliography}

\end{document}